\documentclass{jfm}

\usepackage{graphicx}
\usepackage{newtxtext}
\usepackage{newtxmath}
\usepackage{natbib}
\usepackage{hyperref}
\hypersetup{
    colorlinks = true,
    urlcolor   = blue,
    citecolor  = black,
}

\usepackage{caption}
\newcommand{\RomanNumeralCaps}[1]
\linenumbers

\usepackage{xcolor}

\title{Capturing the Edge of Chaos as a Spectral Submanifold in Pipe Flows }

\author{Bálint Kaszás\aff{1}
  \and George Haller\aff{1} \corresp{\email{georgehaller@ethz.ch}}}

\affiliation{\aff{1}Institute for Mechanical Systems, ETH Z\"{u}rich Leonhardstrasse 21, 8092 Z\"{u}rich, Switzerland}

\begin{document}
\maketitle

\begin{abstract}
An extended turbulent state can coexist with the stable laminar state in pipe flows. We focus here on short pipes with additional discrete symmetries imposed. In this case, the boundary between the coexisting basins of attraction, often called the edge of chaos, is the stable manifold of an edge state, which is a lower-branch traveling wave solution. We show that a low-dimensional submanifold of the edge of chaos can be constructed from velocity data using the recently developed theory of spectral submanifolds (SSMs). These manifolds are the unique smoothest nonlinear continuations of nonresonant spectral subspaces of the linearized system at stationary states. Using very low dimensional SSM-based reduced-order models, we predict transitions to turbulence or laminarization for velocity fields near the edge of chaos.

\end{abstract}



\section{Introduction}
\label{sec:intro}
Flow transition to turbulence has been a topic of great interest for a long time (\cite{eckhardt_TurbulenceTransitionPipe2008}). As the laminar state remains stable for all Reynolds numbers in plane Couette flow and pipe flow (see \cite{schmid_StabilityTransitionShear2001, meseguer_LinearizedPipeFlow2003}), the transition to turbulence is not triggered by a linear instability in these flows. This necessitates the use of inherently nonlinear methods for the analysis of transitions in such flows.


Thanks to the increase in computational power, it is now possible to explore structures directly in the phase space of the Navier-Stokes equations. In this fashion, special invariant solutions that govern the flow behavior have been identified as fixed points or periodic orbits.
Starting with the discovery of the upper and lower-branch fixed points of plane Couette flow by \cite{nagata_ThreedimensionalFiniteamplitudeSolutions1990}, an extensive library of Exact Coherent States (ECS) has been assembled in various flow configurations (\cite{waleffe_ExactCoherentStructures2001, graham_ExactCoherentStates2021}), many of which have also been observed experimentally (\cite{hof_ExperimentalObservationNonlinear2004}).
An important phenomenon revealed by these dynamical system approaches is that the laminar state could coexist with the turbulent one in pipe flows (\cite{duguet_TransitionPipeFlow2008}). For a discussion on the emergence of sustained turbulence, we refer to \cite{avila_OnsetTurbulencePipe2011a} and the recent review by \cite{avila_review_2023}.


We focus here on flow in a periodic domain, where turbulence is not sustained (\cite{hof_RepellerAttractorSelecting2008}) and hence can be characterized as a chaotic saddle (\cite{brosa_TurbulenceStrangeAttractor1989, lai_TransientChaos2011}). This is also true for the dynamics of individual puffs, whose lifetimes grow rapidly with the Reynolds number (\cite{avila_TransientNatureLocalized2010}). Therefore, even if turbulence has a finite lifetime, this lifetime can often be greater than all practically relevant time scales (\cite{lai_GeometricPropertiesChaotic1995}). This makes it feasible to define the boundary that separates trajectories immediately converging to the laminar state from those exhibiting transiently chaotic dynamics as the {\em edge manifold} or {\em the edge of chaos} (\cite{skufca_EdgeChaosParallel2006, schneider_TurbulenceTransitionEdge2007}). Embedded within the edge of chaos, saddle-type {\em edge states} exist (\cite{skufca_EdgeChaosParallel2006, kerswell_RecurrenceTravellingWaves2007, delozar_EdgeStatePipe2012}) whose stable manifolds act as the boundary between the two types of behaviors.  \cite{schneider_TurbulenceTransitionEdge2007} and \cite{mellibovsky_2009} found that the dynamics within the edge may be chaotic. 

Specifically, in short pipes under appropriate symmetry restrictions, the edge of chaos is formed as the stable manifold of a traveling wave solution, known as the lower-branch (LB) traveling wave (see \cite{pringle_AsymmetricHelicalMirrorSymmetric2007, duguet_TransitionPipeFlow2008, kerswell_RecurrenceTravellingWaves2007}). In this case, the edge manifold is a smooth codimension-one invariant manifold that guides trajectories towards the lower-branch traveling wave. The edge manifold and the edge state play a crucial role in the transition to turbulence and the decay from turbulence, as discussed by \cite{delozar_EdgeStatePipe2012}.

The most reliable way to probe the edge has been the edge tracking algorithm introduced by \cite{itano_DynamicsBurstingProcess2001} and \cite{skufca_EdgeChaosParallel2006} in which one selects a turbulent trajectory and a laminarizing trajectory and bisects them. Based on the behavior of the resulting trajectory (whether it is turbulent or laminarizing) one then takes a new bisection. This process is repeated until a trajectory is found that is neither turbulent nor laminarizing for long times (see also \cite{beneitez_EdgeTrackingSpatially2019}).
The bisection method can be aided by stabilizing the edge states, as discussed by \cite{willis_2017}. They propose a simple feedback control scheme, based on the Reynolds number, to remove the unstable directions of the edge state. Forward integration of the controlled system thus results in rapid convergence towards the edge. \cite{linkmann_knierim_zammert_eckhardt_2020} have adapted this method to prevent the control scheme from introducing additional unstable directions.
An alternative characterization of the edge of chaos is given by \cite{beneitez_EdgeManifoldLagrangian2020}, who reinterpret the edge as a Lagrangian coherent structure (see \cite{haller_LagrangianCoherentStructures2015, haller_TransportBarriersCoherent2023}). 

Reduced-order models promise an efficient way to describe the transition to turbulence. They are typically obtained from a Galerkin projection of the Navier-Stokes equations onto a small set of spatial modes (\cite{eckhardt_TransitionTurbulenceShear1999, moehlis_LowdimensionalModelTurbulent2004, joglekar_GeometryEdgeChaos2015}). Alternatively, data-driven methods can infer reduced-order models directly from time-resolved simulations or experiments. 

Common approaches to data-driven reduced-order modeling are linear, such as the dynamic mode decomposition introduced by \cite{schmid_DynamicModeDecomposition2010} or the Koopman-mode expansion (\cite{rowley_SpectralAnalysisNonlinear2009}). Such linear methods cannot capture the characteristically nonlinear bistability of shear flows, as was demonstrated in detail by \cite{page_KoopmanModeExpansions2019}. However, one can build formal nonlinear models as expansions based on linear models. For example, \cite{ducimetiere_WeakNonlinearityStrong2022} used the spectrum of the resolvent operator and multiple-scale expansion to derive Stuart-Landau-type amplitude equations (\cite{landau_FluidMechanics1959}) for flows exhibiting non-normality. Among other examples, the amplitude equations were then used to predict the energy of the response observed in a plane Poiseuille flow subjected to harmonic forcing.


In contrast to approximate linear models, invariant manifold based methods provide a mathematically rigorous foundation for reduced-order models that capture nonlinear features. An early demonstration of this fact was the approximate inertial manifold approach, which argues that the dynamics evolves on a finite, but still high-dimensional invariant manifold (\cite{foias_ComputationInertialManifolds1988}). Recently, deep learning methods have shown promise in learning the dynamics within the inertial manifold (see \cite{linot_DeepLearningDiscover2020}). Despite all these advances, however, inertial manifolds have not been proven to exist in Navier-Stokes flows. 

We focus here on the recently introduced spectral submanifolds (SSMs). These are low-dimensional invariant manifolds connected to known stationary states, to which the dynamics of a (possibly infinite dimensional) dynamical system can be reduced (\cite{haller_NonlinearNormalModes2016a, kogelbauer_RigorousModelReduction2018}). SSMs are the smoothest nonlinear continuations of spectral subspaces of the linearized system in a neighborhood of a stationary state, such as a fixed point or a periodic orbit.
Based on earlier work by \cite{cabre_ParameterizationMethodInvariant2003}, SSM-based model reduction exploits the existence and uniqueness of spectral submanifolds, both of which are guaranteed when the spectrum of the linearized system within the spectral subspace is not in resonance with the rest of the spectrum outside that subspace. Restricting the dynamics to SSMs associated with the slowest spectral subspace then provides a mathematically exact reduced-order model. In some cases, such a slow SSM may contain the global attractor of the system and can be considered an inertial manifold.

In its initial formulation, SSM reduction constructed the parametrization of the invariant manifold and the reduced dynamics as a Taylor-expansion around a steady state, which proved fruitful in reduced-order modeling for general mechanical systems (e.g., \cite{jain_HowComputeInvariant2022, li_ModelReductionConstrained2023}).
In addition to its strict mathematical foundation, a noteworthy advantage of SSM-based model reduction over projection-based methods is that the dimension of the slowest nonresonant spectral subspace  a priori determines the dimension of the reduced-order model. Therefore, one can simply increase the order of the Taylor expansion aprroximating the SSM and its reduced dynamics, without increasing the dimension of the reduced model to achieve higher accuracy. 

By now, \cite{buza_SpectralSubmanifoldsNavierStokes2023} has also established the existence of certain classes of SSMs for the Navier-Stokes equations. In a recent development, \cite{haller_NonlinearModelReduction2023} introduced generalized families of (secondary) SSMs that can have either a lower degree of smoothness (fractional SSMs) or can be tangent to a spectral subspace containing stable and unstable modes at the same time (mixed-mode SSMs). In the present study, we will invoke these results to construct mixed-mode SSMs as a basis for reduced-order models for transitions in a pipe flow. We rely on recent work by \cite{cenedese_DatadrivenModelingPrediction2022} that combines SSM theory with data-driven methods to construct reduced-order models directly from experimental or numerical data. We will use an implementation of these results in the open-source package \texttt{SSMLearn} (\cite{cenedese_SSMLearn2021}), which has already been successfully applied to fluids problems, such as sloshing  in a horizontally forced tank (\cite{axas_FastDatadrivenModel2023}).

Using the same data-driven method, \cite{kaszas_DynamicsbasedMachineLearning2022} showed that an SSM-based model accurately predicts the time evolution of individual trajectories in the phase space of plane Couette flow. While indeed yielding accurate and predictive models, those results only covered transitions between nonchaotic states, such as fixed points and periodic orbits.
In this contribution, we show that similar results can be applied even when the phase space contains turbulent behavior supported on a chaotic saddle.

In particular, we target the slowest two-dimensional SSM of the edge state to characterize the dynamics of laminarization and transition to turbulence. We use data generated by the open-source solver \texttt{Openpipeflow} of \cite{willis_OpenpipeflowNavierStokes2017}. We do not seek to capture any turbulent dynamics because we have no accurate way to approximate the dimension of the underlying chaotic saddle and hence cannot guarantee that it will be contained in an SSM. Instead, we show that a data-driven SSM-reduced model accurately captures the slowest submanifold in the edge of chaos, thus yielding an explicit parametrization of the most influential dynamical structures within the edge manifold. We have also made the data and the code supporting the analysis available under the repository \cite{cenedese_SSMLearnPy2023} as well as in the form of \texttt{JFM Notebooks}.

\section{Setup}

We consider the incompressible Navier-Stokes equations for the velocity field $\mathbf{u}$ and the pressure $p$ in a domain $\Omega \subset \mathbb{R}^3$, given by
\begin{equation}
    \label{ns}
    \frac{\partial \mathbf u}{\partial t} + \left ( \mathbf{u}\nabla\right ) \mathbf{u} = -\nabla p + \frac{1}{\text{Re}}\Delta \mathbf{u} + \mathbf{q}
    \quad \nabla \cdot \mathbf{u} = 0,
\end{equation}
where $\mathbf{q}$ is a body force needed to sustain a constant mass flux. The domain is taken as a circular pipe, 
\begin{align}
    \label{domains}
    \Omega_{\text{pipe}}&=\left\{(r, \varphi, z)\in \mathbb{R}^3 | \ 0 \leq r\leq R, \varphi \in [0, 2\pi), 0\leq z\leq L\right\},
\end{align}
defined in cylindrical coordinates with $x=r\cos \varphi$, $y=r\sin \varphi$. The Reynolds number is defined as $\text{Re}=\frac{R U_{cl.}}{\nu}$, where $U_{cl.}$ is the center line velocity of the laminar state and $\nu$ is the kinematic viscosity. Equation \eqref{ns} is nondimensionalized by the pipe radius $R$ and the center line velocity $U_{cl}$. The laminar state is the Hagen-Poiseuille flow, given by
\begin{equation}
\label{eq:pois}
    \mathbf{u}_{HP}(r) = \begin{pmatrix} 0 \\ 0 \\ 1-r^2 \end{pmatrix}.
\end{equation}
The domain is assumed to be periodic in the $z$ direction, so that $~\mathbf{u}(r,\varphi,z,t) = \mathbf{u}(r,\varphi,z+L,t)$. Following previous studies (\cite{duguet_TransitionPipeFlow2008, willis_RevealingStateSpace2013}), we approximate the velocity field as a truncated Fourier expansion up to order $M$ in the azimuthal direction and up to order $K$ in the streamwise direction as
\begin{equation}
    \mathbf{u}(r,\varphi, z, t) = \mathbf{u}_{HP} + \sum_{m=-M}^{M} \sum_{k=0}^K \mathbf{A}_{mk}(r,t)e^{im_p m \varphi} e^{ik\alpha z},
    \label{eq:fourier}
\end{equation}
where $\mathbf{A}_{mk}(r,t)$ is a 3-vector of Fourier-amplitudes, $m_p$ determines the fundamental period in the angular direction and $\alpha = \frac{2\pi}{L}$. Throughout this study, we fix the number of modes used as $M=16$, $K=16$, and fix $m_p=2$ and $\alpha = 1.25$. The latter sets the length of the pipe as $L \approx 5.02$, which corresponds to a minimal flow unit studied in \cite{willis_RevealingStateSpace2013, willis_2017}. Our spatial resolution also corresponds to that of \cite{willis_RevealingStateSpace2013}, who have confirmed that the dynamics are sufficiently resolved by this level of discretization.

We use \texttt{Openpipeflow} (\cite{willis_OpenpipeflowNavierStokes2017}) to perform the discretization \eqref{eq:fourier} with an additional finite-difference approximation at $64$ points for the radial derivative. The discretized Navier-Stokes equations are then integrated using a pressure Poisson equation formulation to enforce incompressibility. The time-resolved Navier-Stokes solutions are viewed as trajectories of a large system of coupled ordinary differential equations, 
\begin{equation}
    \label{eq:ns_dynsys}
    \frac{d }{dt}\mathbf{x} = \mathcal{N}(\mathbf{x}), \quad \mathbf{x}\in \mathbb{R}^{k},
\end{equation} 
where $\mathbf{x}$ is the vector comprised of the discretized degrees of freedom with $k\in O(10^5)$ and $\mathcal{N}$ is the generator of the time evolution. As expressed by \eqref{eq:fourier}, the solver returns the deviations with respect to the laminar state, such that the origin $\mathbf{x}=0$ corresponds to \eqref{eq:pois}. Distinguished solutions of this dynamical system are ECSs discussed in the Introduction. \cite{budanur_RelativePeriodicOrbits2017} has shown that some of these ECSs, unstable (relative) periodic orbits, are indeed central organizers of turbulence. 

  \begin{figure}
     \includegraphics[width = \textwidth]{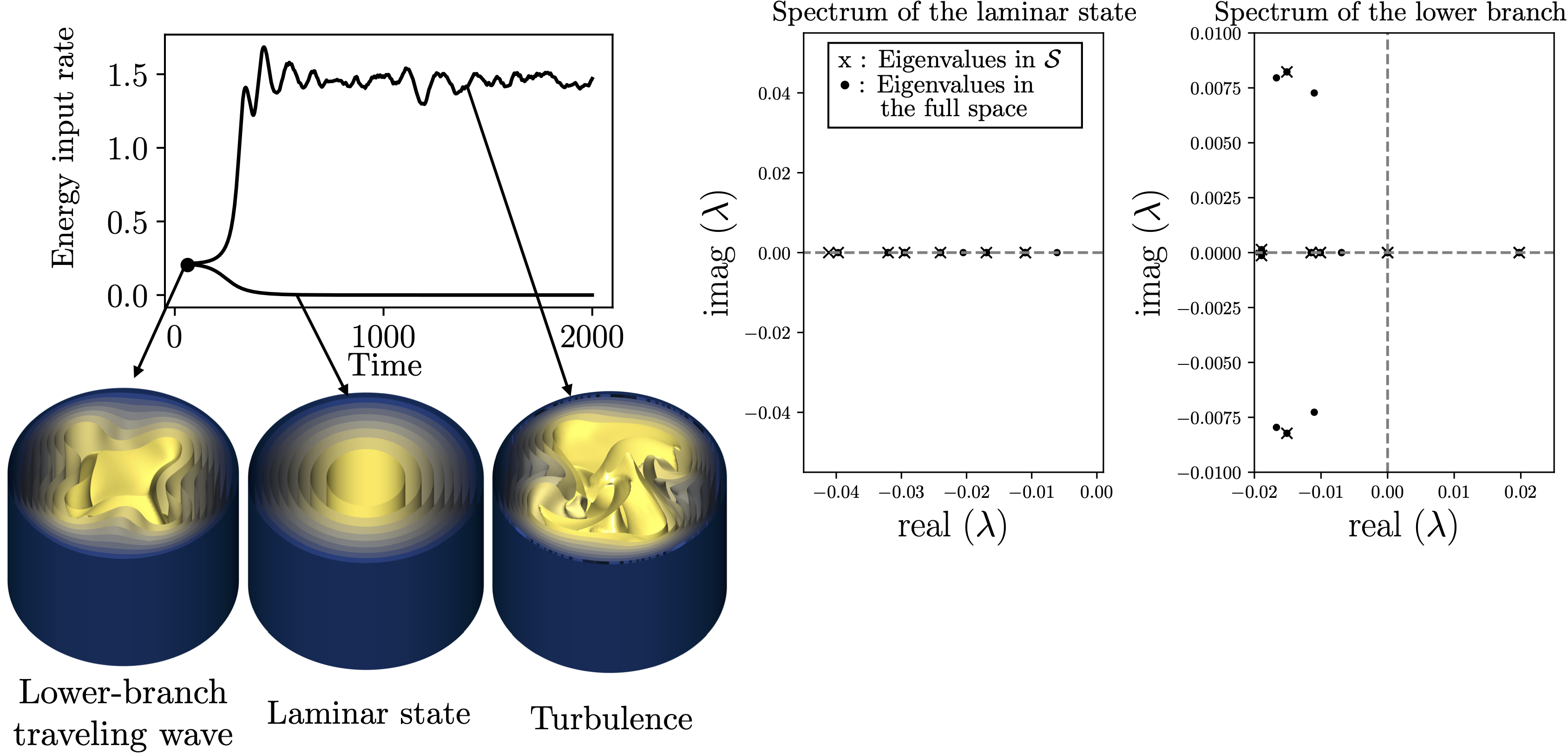}     
     \caption{Left panel: Time evolution of the energy input rate of two trajectories started near the lower-branch traveling wave. Below we illustrate the flow fields corresponding to different snapshots during the time evolution by showing isosurfaces of the streamwise velocity $u_z$. Middle and right panels: Eigenvalue configuration of the laminar state and the lower-branch traveling wave, respectively. Crosses denote the eigenvalues in the $\mathcal{S}$-invariant subspace discussed in the text and circles denote the full-space eigenvalues. The parameters are chosen as $\text{Re}=2,400; \alpha = 1.25$. }
     \label{fig:fig1}
 \end{figure}

For low Reynolds numbers, the only stable state
is the laminar state, while for transitional flows above $\text{Re}\sim 2,000$, the phase space of \eqref{eq:ns_dynsys} is divided into two domains with characteristically different behaviors. 
Some initial conditions immediately return to the vicinity of the state \eqref{eq:pois}, while others transition to turbulence. 

The left panel of Fig. \ref{fig:fig1} shows two trajectories of \eqref{eq:ns_dynsys} with distinct dynamic behaviors, as seen in the time evolution of their normalized energy input rate $I$, defined as
\begin{equation}
    I' = \frac{1}{V} \oint_{\partial \Omega_{\text{pipe}}}   d\mathbf{S}\cdot \mathbf{u}p , \quad I = \frac{I'}{I'_{HP}} - 1,
\end{equation}
with the volume of the domain $\Omega_{\text{pipe}}$ denoted as $V$. The energy input rate $I'$ measures the external power supplied to the system to satisfy the constant mass flux condition. Similarly, the normalized rate of energy dissipation is given by
\begin{equation}
    D' = \frac{1}{\text{Re}} \left \Vert \text{curl }\mathbf{u} \right \Vert^2 , \quad D = \frac{D'}{D'_{HP}}-1,
\end{equation}
with the norm of a vector field defined as
\begin{equation}
\label{eq:normdef}
\left \Vert \mathbf{f} \right \Vert := \sqrt{\langle \mathbf{f}, \mathbf{f} \rangle} = \left( \frac{1}{V}\int_{\Omega_{\text{pipe}}}dV \ \mathbf{f}\cdot \mathbf{f}\right)^{\frac{1}{2}}.
\end{equation}
In the above definitions we normalize the energy input and dissipation values by their laminar values and subtract $1$ so that the laminar state corresponds to $I_{HP}=0$, $D_{HP}=0$.
An energy balance can be derived from the inner product $\langle \mathbf{u}, \frac{\partial \mathbf{u}}{\partial t}\rangle$ using \eqref{ns} (see \cite{waleffe_2011_w}). The energy input rate, the dissipation rate, and the kinetic energy, defined as $E = \Vert \mathbf{u}\Vert^2/2$, satisfy
\begin{equation}
    \dot{E} = I' - D'. 
\end{equation}

\cite{skufca_EdgeChaosParallel2006} discovered that a codimension-one surface, the {\em edge manifold} locally acts as a barrier that separates laminarizing and turbulent trajectories. This manifold is the stable manifold of an unstable traveling wave solution known as the lower-branch discussed by \cite{kerswell_RecurrenceTravellingWaves2007}, \cite{willis_RevealingStateSpace2013} and  \cite{budanur_ComplexityLaminarturbulentBoundary2018}. This solution is invariant with respect to the shift-reflect symmetry
\begin{equation}
\mathcal{S}\begin{pmatrix}
    u_r(r,\varphi, z) \\ u_\varphi(r,\varphi, z) \\ u_z(r,\varphi, z) 
\end{pmatrix} = \begin{pmatrix}
    u_r\left(r,\varphi, z-\frac{L}{2}\right) \\ -u_\varphi\left(r,\varphi, z-\frac{L}{2}\right) \\ u_z\left(r,\varphi, z-\frac{L}{2}\right)
\end{pmatrix}.
\end{equation}
For convenience, we perform the calculations in the $S-$invariant subspace as is customary in the literature (\cite{willis_RevealingStateSpace2013, budanur_ComplexityLaminarturbulentBoundary2018}). 

Spectral submanifolds (SSMs, \cite{haller_NonlinearNormalModes2016a}) have emerged as useful tools for reduced-order modeling of large-scale systems. They are defined as the unique invariant manifolds of \eqref{eq:ns_dynsys} that are the smoothest continuations of the invariant spectral subspaces of the linearization of \eqref{eq:ns_dynsys} at hyperbolic fixed points or periodic orbits. Although the existence conditions for SSMs in infinite-dimensional systems are difficult to verify (see \cite{kogelbauer_RigorousModelReduction2018}), this analysis has been carried out by \cite{buza_SpectralSubmanifoldsNavierStokes2023} for the Navier-Stokes equations. 

As any numerical solution of system \eqref{ns} is inevitably finite-dimensional, invoking the finite-dimensional results of \cite{haller_NonlinearNormalModes2016a} is sufficient in our present setting. These results  guarantee that a stable hyperbolic fixed point has a hierarchy of spectral submanifolds attached to it provided that the spectral non-resonance conditions mentioned in the Introduction are met. 

A notable hyperbolic anchor point for SSM construction is the laminar Hagen-Poiseuille flow \eqref{eq:pois}. The spectrum of the linearized Navier-Stokes equation \eqref{ns} at \eqref{eq:pois} is discussed by  \cite{schmid_StabilityTransitionShear2001}. 
The slowest spectral subspace inferred from these calculations corresponds to the least stable  real eigenvalue or complex conjugate eigenvalue pair. For our geometry with $\alpha=1.25$ and $m_p=2$, these are the streamwise-independent modes with $k=0$ in \eqref{eq:fourier}, which have the form
\begin{equation}
    \label{eq:bessel}
    \mathbf{v}_{nm'}(r,\varphi) = \begin{pmatrix}
        0 \\ 0 \\ J_{m'}(\sqrt{-\lambda_{n,m'}\text{Re}}r)
    \end{pmatrix}e^{i m' \varphi }.
\end{equation}
Here, $m'=m_pm$ and $J_{m}$ is the $m-$th Bessel function of the first kind. The corresponding eigenvalue is real and is given by
$$
\lambda_{n,m'} = -\frac{j^2_{n,m'}}{\text{Re}},
$$
where $j_{n,m'}$ denotes the $n-$th root of $J_{m'}$. Thus, the slowest spectral subspace is one-dimensional and is obtained for $n=1, m'=m_p = 2$.

The nonlinear term $(\mathbf{u}\nabla)\mathbf{u}$ vanishes identically along the modes \eqref{eq:bessel}, which means that the spectral subspace spanned by \eqref{eq:bessel} is also an invariant manifold of the nonlinear system \eqref{ns}. By the uniqueness of the  slowest analytic spectral submanifold, the spectral subspace spanned by \eqref{eq:bessel} is necessarily the unique, analytic spectral submanifold of system \eqref{ns} anchored at the state \eqref{eq:pois}.
As such, this SSM cannot carry nonlinear dynamics. This conclusion is also supported by the results of \cite{joseph_ContributionsNonlinearTheory1971}, who show that a streamwise-independent flow such as \eqref{eq:bessel} must ultimately decay. \cite{kogelbauer_NotesSpectralSubmanifolds2020} also investigated slow SSMs of the laminar state in pipe flow for a range of parameter values and found no nontrivial behavior in Taylor approximations of those SSMs. Indeed, \cite{haller_NonlinearModelReduction2023} find that heteroclinic connections among steady states generally occur along invariant manifolds of finite smoothness. Their results indicate that typical SSMs in the flow are only once continuously differentiable at the laminar state. This can be deduced from the spectrum of the laminar state (see Fig. \ref{fig:fig1} middle panel).

The other anchor point candidate for SSM-based model order reduction is the lower-branch traveling wave, as seen from the spectrum of the linearization of \eqref{ns} at that solution in the right panel of Fig. \ref{fig:fig1}. 

\subsection{Symmetry reduced phase space}
The Navier-Stokes equations \eqref{ns} have two continuous symmetries. Streamwise translations $\sigma_l$ and rotations $\sigma_\theta$. This means that any solution $\mathbf{u}(t)$ is physically equivalent to all solutions along orbits of the group generated by the two shifts, denoted as $\sigma_l$ and $\sigma_\theta$, that is to the set

\begin{equation}
\label{eq:grouporbit}
\Sigma(\mathbf{u}(t)) = \left\{\sigma_l \sigma_\theta \mathbf{u}(r,\varphi, z, t)= \mathbf{u}(r,\varphi-\theta, z-l, t) \vert \theta \in [0, 2\pi), l \in [0, L)  \right\}
\end{equation}

Imposing the shift-reflect symmetry $\mathcal{S}$ (i.e., restricting \eqref{ns} to the $\mathcal{S}-$invariant subspace) eliminates one of these shifts, but solutions in the invariant subspace can still be translated freely in the streamwise direction. This translational invariance is reflected by the appearance of an eigenvector with zero eigenvalue for the lower-branch traveling wave.

This zero eigenvalue formally renders the lower-branch traveling wave a non-hyperbolic fixed point and prevents us from concluding the existence of SSMs anchored at this fixed point. However, since non-hyperbolicity arises here due to a continuous symmetry, we can use the {\em method of slices} (\cite{froehlich_ReductionContinuousSymmetries2012}) to eliminate this symmetry. This method factorizes the phase space of \eqref{eq:ns_dynsys} by establishing an equivalence relation along group orbits. From each group orbit \eqref{eq:grouporbit}, we select the single trajectory that is closest to the lower-branch traveling wave in the norm \eqref{eq:normdef}. This construction has already been used to study the same system by \cite{willis_RevealingStateSpace2013} and to aid dynamic mode decomposition (\cite{marensi_SymmetryreducedDynamicMode2023}).

The method of slices then yields a dynamical system that has a lower dimension than the original one. The dynamics along the group orbit, which we may denote by a phase-type variable $\psi$, can also be recovered. Thus, formally, the dynamics is governed by 
\begin{align}
    \dot{\mathbf{x}}_R &= \mathcal{N}_R (\mathbf{x}_R), \quad \mathbf{x}_R\in \mathbb{R}^{k-1} \label{eq:symmetry_reduced_phasespace}\\
    \dot{\psi} &= f(\mathbf{x}_R, \psi). 
\end{align}

In the following, we work with the $\mathbf{x}_R$ component of the dynamical system obtained from the method of slices which makes the phase space $d=k-1$ dimensional. We illustrate the trajectories computed in the full space in the upper left panel of Fig. \ref{fig:fig2}. It can be seen that trajectories evolve in spirals, which is a signal of their time evolution along the group orbit. 

The method of slices eliminates this behavior, as can be seen in the right panel of Fig. \ref{fig:fig2}. 
Note also that physically inherent quantities, such as the energy input rate and the dissipation are the same along all points of the group orbit, which can be seen in the left panel of Fig. \ref{fig:fig2}. Eliminating the symmetry thus does not change the physically relevant observables of a trajectory, since macroscopic quantities must be the same for flow fields related to each other by a symmetry transformation. 

\section{Results}
\subsection{Spectral submanifolds of the lower-branch traveling wave}

In the symmetry reduced phase space, the lower-branch traveling wave becomes a hyperbolic fixed point with a single unstable eigenvalue. By the classic unstable manifold theorem, there exists a one-dimensional unstable manifold tangent to the unstable eigenvector and a codimension-one stable manifold. 
The unstable manifold of the lower-branch traveling wave forms a heteroclinic connection with the laminar state. \cite{kaszas_DynamicsbasedMachineLearning2022} demonstrated that such heteroclinic orbits can serve as one-dimensional reduced-order models, as also supported by the experiments of \cite{suri_ForecastingFluidFlows2017}. 

To capture more of the transient dynamics near the heteroclinic orbit and obtain an approximation for the basin boundary of the laminar state, we extract here a two-dimensional invariant manifold that can intersect the edge. In order to construct an attracting manifold, we need to take the slowest two-dimensional SSM of the lower-branch traveling wave. This SSM is tangent to both its single unstable eigenvector and its slowest stable eigenvector and hence contains both the heteroclinic orbit and the slowest submanifold of the edge of chaos.

Such a manifold is often called a pseudo-unstable manifold ( \cite{delallave_IrwinProofPseudostable1995}), whose existence can also be concluded from the recent results of \cite{haller_NonlinearModelReduction2023} or \cite{buza_SpectralSubmanifoldsNavierStokes2023}. \cite{haller_NonlinearModelReduction2023} also find a large variety of generalized spectral submanifolds of \eqref{eq:ns_dynsys}, including ones of limited smoothness (fractional SSMs) and others of mixed stability type (mixed-mode SSMs). These results apply in finite dimensions and give $C^{\infty}$-smooth, mixed-mode SSMs under the same nonresonance conditions as those of \cite{sternberg_StructureLocalHomeomorphisms1958}.

As discussed by \cite{haller_NonlinearModelReduction2023}, exact resonances among complex eigenvalues are unlikely in a typical finite-dimensional system, such as the numerical discretization \eqref{eq:ns_dynsys}. We thus expect that a unique, $C^\infty$, mixed-mode SSM of the lower-branch traveling wave exists. 
We list the corresponding nonresonance conditions in the Appendix and verify that they hold for a subset of the spectrum. 
Alternatively, one may invoke the results of \cite{buza_SpectralSubmanifoldsNavierStokes2023}, who showed that a $C^1$-smooth pseudo-unstable manifold exists for the Navier-Stokes equations \eqref{ns}.

Based on these recent technical developments, we can apply the data-driven methodology of \cite{cenedese_DatadrivenModelingPrediction2022} to discover this mixed-mode SSM from simulation data.
We follow the approach of \cite{kaszas_DynamicsbasedMachineLearning2022} and introduce the square roots of the variables $I$ and $D$ as $J := \sqrt{|I|}$ and $K := \sqrt{|D|}$. This is necessary, because the velocity field $\mathbf{u}$ is a non-differentiable function of $I$ and $D$ at the origin, due to their quadratic dependence on the components of $\mathbf{u}$. Thus, we parametrize the spectral submanifolds with the variables $J$ and $K$. Specifically, we seek a two-dimensional invariant manifold in the phase space of \eqref{eq:symmetry_reduced_phasespace} of the form
\begin{equation}
\label{eq:reduced_mfd}
\mathbf{x}_R =  \sum_{1\leq n + m \leq M_p} \mathbf{c}_{nm}K^mJ^n, \quad \mathbf{c}_{nm}\in \mathbb{R}^d.
\end{equation}
The polynomial-type dependence up to order $M_p$ on the reduced coordinates in \eqref{eq:reduced_mfd} is justified because polynomials are universal approximators (\cite{rudin_PrinciplesMathematicalAnalysis1976}). It is also motivated by the success of local Taylor-approximations used in the original methods (\cite{haller_NonlinearNormalModes2016a}) for SSM-based model order reduction. 

To determine the coefficient vectors $\mathbf{c}_{nm}$ in \eqref{eq:reduced_mfd} we initialize training trajectories that lie on the mixed-mode SSM of the lower-branch traveling wave. 
The parametrization of the manifold in physical coordinates can then be recovered from \eqref{eq:reduced_mfd} using the discretization \eqref{eq:fourier} performed by \texttt{Openpipeflow} (\cite{willis_OpenpipeflowNavierStokes2017}). This results in 
\begin{equation}
\label{eq:mfd}
    \mathbf{u}(r,\varphi,z) = \mathbf{W}(J,K) = \mathbf{u}_{HP}(r) + \sum_{1\leq n+m\leq M_p}\mathbf{w}_{nm}(r, \varphi,z)K^mJ^{n},
\end{equation}
where the coefficient-functions $\mathbf{w}_{nm}(r,\varphi,z)$ are determined by $\mathbf{c}_{nm}$.

To generate initial conditions, we start with the lower-branch traveling wave discussed by \cite{kerswell_RecurrenceTravellingWaves2007} at $Re = 2,400$. The traveling wave is found using a Newton-Krylov scheme (\cite{viswanath_RecurrentMotionsPlane2007}). This method also returns the leading eigenvalues and eigenvectors of the traveling wave (computed in a co-moving frame) (see \cite{willis_EquilibriaPeriodicOrbits2019}). 

For training, we take a total of six trajectories, four of which were randomly initialized at a distance of $10^{-4}$ from the lower-branch traveling wave. The distance is measured as the Euclidean distance in the phase space of \eqref{eq:symmetry_reduced_phasespace}. In this metric, the distance between the laminar state and the lower-branch traveling wave is 0.72. We reserve an additional randomly initialized trajectory for validation. After the initial transients, the trajectories approach the unstable manifold of the lower-branch traveling wave. To also capture the dynamics in the slowest stable direction, we construct two more training trajectories that are approximately constrained to the edge. We use the edge-tracking algorithm of \cite{itano_DynamicsBurstingProcess2001} started from initial guesses lying along the subspace spanned by the slowest stable eigenvector. The energy input and dissipation variables are computed using \texttt{Openpipeflow} from the spectral representation of the flow fields. 

\begin{figure}
 \centering
     \includegraphics[width = 0.9\textwidth]{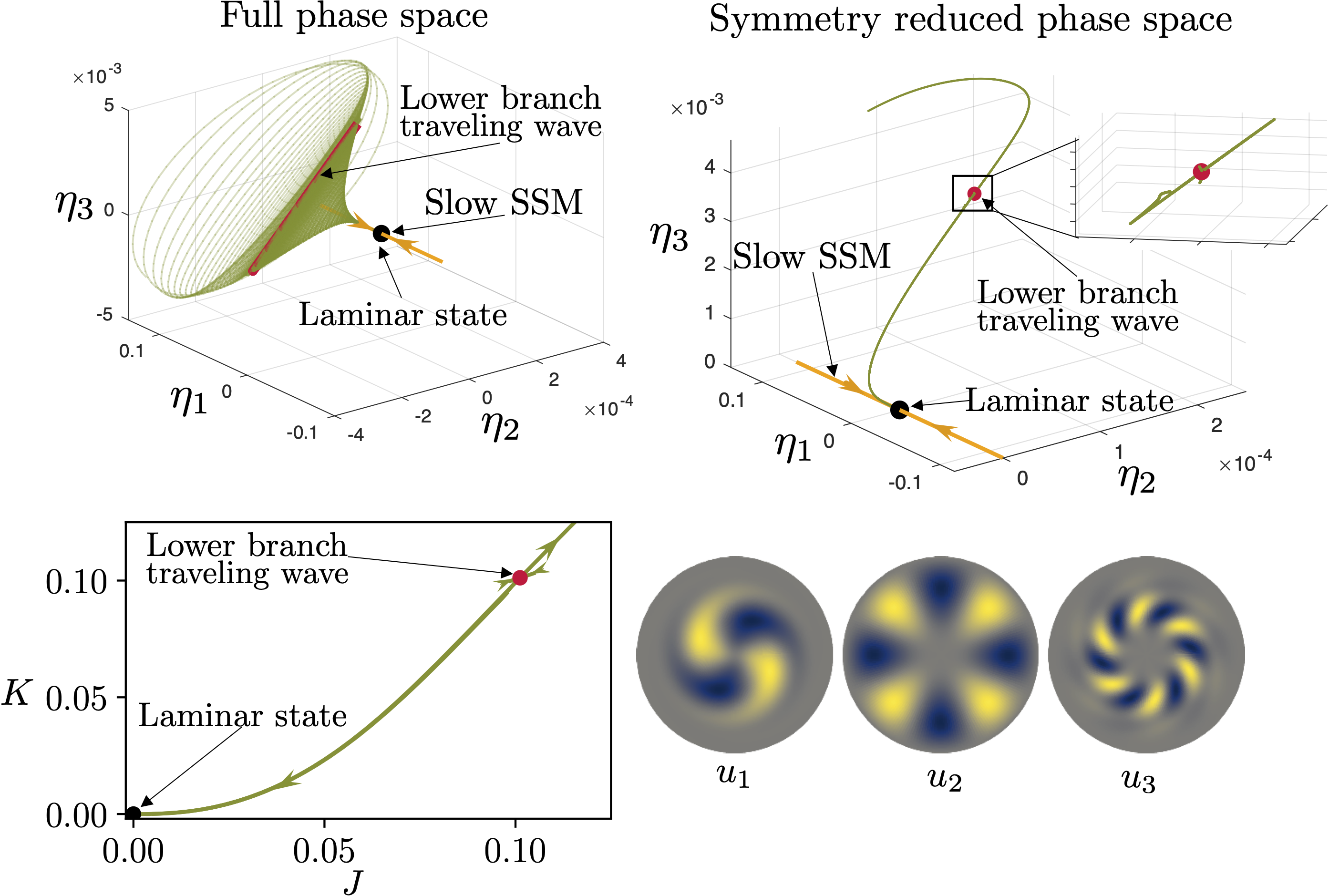}
     \caption{Upper left panel: Trajectories initialized near the lower-branch traveling wave displayed in the space spanned by projections onto the three most dominant spatial modes of the lower-branch. The coordinates are defined as $\eta_i = \langle \mathbf{u}, u_i\rangle$, where the three dominant modes are illustrated in the lower right panel.
    Right panel: projections of the symmetry reduced trajectories onto the same spatial modes. 
     Lower left panel: the same trajectories displayed in the space spanned by $J=\sqrt{I}$ and $K=\sqrt{D}$. The directory, including the data and the Jupyter notebook that
generated this figure can be accessed at \url{https://cocalc.com/share/public_paths/2af692bffb0397d126563ec78c61bb65205d1211}}
     \label{fig:fig2}
 \end{figure}

We show the set of training trajectories in Fig. \ref{fig:fig2}. The upper left panel shows the original trajectories as a projection onto the three most dominant spatial modes, which are displayed in the lower right panel of the figure. In all further computations, we work with the symmetry-reduced phase space and pre-process the trajectories using the method of slices. The symmetry-reduced trajectories can be seen in the upper right panel of Fig. \ref{fig:fig2}. We also show the reduced coordinates, $J=\sqrt{|I|}$ and $K=\sqrt{|D|}$, which coincide for the full-space trajectories and the symmetry-reduced trajectories. 

Note that the method of slices generally can only be applied over a bounded domain. When trajectories cross the border of the chart associated to the template (\cite{froehlich_ReductionContinuousSymmetries2012}), finite jumps are observed in the symmetry-reduced phase space. Figure \ref{fig:fig2} indicates no such singularities in our training trajectories, which prompts us to represent them by a single chart.  \cite{willis_RevealingStateSpace2013} found that a global atlas defined using multiple template states was necessary to construct the symmetry reduced representation of the turbulent state. Since our training trajectories avoid the turbulent state and remain in the neighborhood of the heteroclinic orbit, using the lower-branch traveling wave as the only template state was sufficient.

Given the training trajectories, we identify the coefficient vectors $\mathbf{c}_{nm}$ via a sparsity promoting ridge regression (\cite{hastie2009elements, brunton_DiscoveringGoverningEquations2016}), minimizing the squared norm of the deviation from these trajectories. Once the SSM geometry is identified, we seek the SSM-reduced dynamics in the form
\begin{equation}
\label{eq:reddynamics}
\frac{d}{dt}\begin{pmatrix}J  \\ K \end{pmatrix} = \mathbf{f}(J,K)= \sum_{1\leq n+m\leq M_r} \begin{pmatrix} R^{(J)}_{nm} K^m J^n \\ R^{(K)}_{nm} K^m J^n\end{pmatrix}. \\
\end{equation}
Here the coefficients $R^{(J)}_{nm}$ and $R^{(K)}_{nm}$ are also determined by ridge regression onto $\dot{J}$ and $\dot{K}$ obtained by finite differencing along the training trajectories. We also introduce a constraint to this optimization, forcing the lower-branch traveling wave and the laminar state to be fixed points of \eqref{eq:reddynamics}, as in \cite{kaszas_DynamicsbasedMachineLearning2022}.

The parameters of the regression are the polynomial orders $M_p$ and $M_r$, as well as the weight of the ridge-type penalty term. For simplicity, we take $M_p=M_r$. This choice is motivated by the usual Taylor-expansion representation of SSMs, where the parametrization and the reduced dynamics are computed up to the same order. The value of the polynomial orders and the ridge-type penalty term are determined by cross-validation on a trajectory initialized in the same way as the training set.

We minimize the overall mean-prediction error on the validation trajectory, defined as 
\begin{equation}
    \label{eq:crossval_error}
    \text{Error} = \sum_{i=0}^{n_t} \left \Vert \mathbf{u}_{true}(t_i) - \mathbf{W} \circ F^{t_i}(J_{true}(0), K_{true}(0))\right \Vert,
\end{equation}
where $F^t$ is the time-$t$ flow-map of the reduced dynamics \eqref{eq:reddynamics} and $n_t$ is the number of time-snapshots available along the validation trajectory. To avoid overfitting and promote simpler models, as long as the value of the error \eqref{eq:crossval_error} is similar, we prioritize lower polynomial orders and larger penalty terms. The results are reported with $M_r=M_p=5$, and the penalty term is chosen as $10^{-5}$. We refer to the \texttt{JFM Notebook} accompanying Fig. \ref{fig:fig4} for further details.

We also note that the approximation of the SSM benefits from a richer training set. The accuracy of the reduced-order model increases if more training trajectories are used for the regression or if the temporal resolution of the trajectories is increased, as explained by \cite{cenedese_DatadrivenModelingPrediction2022}.

The lower-branch traveling wave is a fixed point of the reduced-order model by construction, that is $\mathbf{f}(J_{LB}, K_{LB}) = \mathbf{0}$. The Jacobian, $D\mathbf{f}(J_{LB}, K_{LB})$, has eigenvalues ${\lambda^{(+)}_{red.} = 0.0199}$, and ${\lambda^{(-)}_{red.} = -0.0107}$, matching the eigenvalues of the lower-branch in the full-order model obtained by Krylov iteration in \texttt{Openpipeflow}, whose values are  ${\lambda^{(+)} = 0.0198}$, and ${\lambda^{(-)} = -0.0010}$.
\subsection{Predictions of the reduced-order model}
\begin{figure}
     \centering
     \includegraphics[width = .75\textwidth]{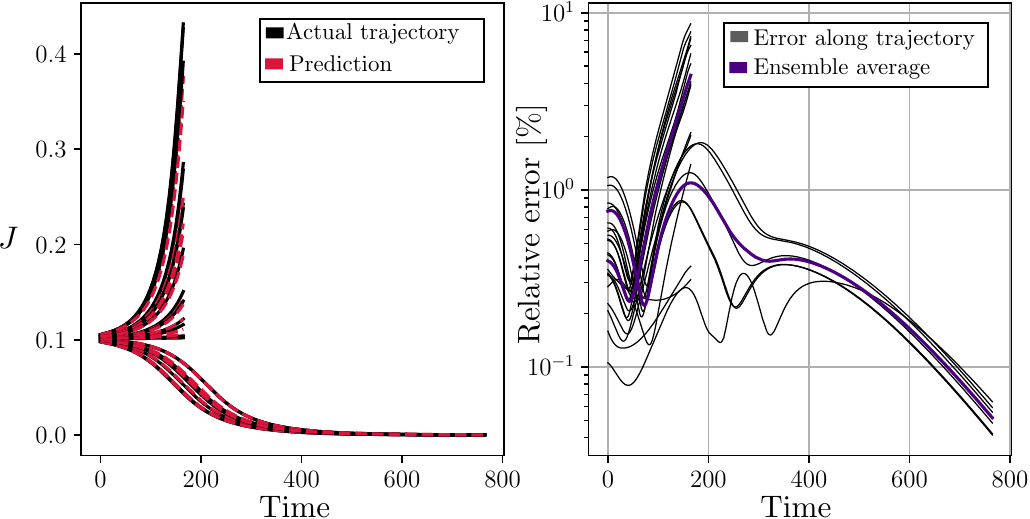}     
     \caption{Predictions of the reduced-order model on an ensemble of 20 test trajectories initialized near the lower-branch traveling wave. In the left panel, the time evolution of their $J$ coordinate is compared to the predictions made by the reduced-order model. For turbulent trajectories, only the first 200 time units of the trajectories are shown.
     In the right panel, the  corresponding relative errors are plotted, defined as the time-dependent quantity $\left \Vert 
     \mathbf{u}_{true}(t_i) - \mathbf{W} \circ F^{t_i}(J_{true}(0), K_{true}(0))\right \Vert/ \text{max }\left \Vert \mathbf{u}_{true}\right \Vert$. Ensemble-averages of the relative error are computed. The errors of turbulent and laminar trajectories are averaged separately. }
     \label{fig:fig3}
 \end{figure}

To assess the predictive power of the two-dimensional, SSM-reduced model obtained above, we also generate an ensemble of test trajectories distributed over a sphere of radius $10^{-4}$ around the lower-branch traveling wave in the phase space. Some of these trajectories transition to turbulence while others laminarize, but we can make predictions based on the reduced-order model for all of them. In Fig. \ref{fig:fig3} we compare the time evolution of the $J$ coordinate of the test trajectories to their predicted counterparts. While the reduced-order model accurately differentiates between turbulent and laminar trajectories, the turbulent trajectories are only reliably modeled for 200 time units, since the model was not trained on trajectories in the turbulent state. 
The right panel of Fig. \ref{fig:fig3} shows the overall relative error of the predictions \eqref{eq:crossval_error} for the ensemble of test trajectories. 

\begin{figure}
   \centering
     \includegraphics[width = 0.68\textwidth]{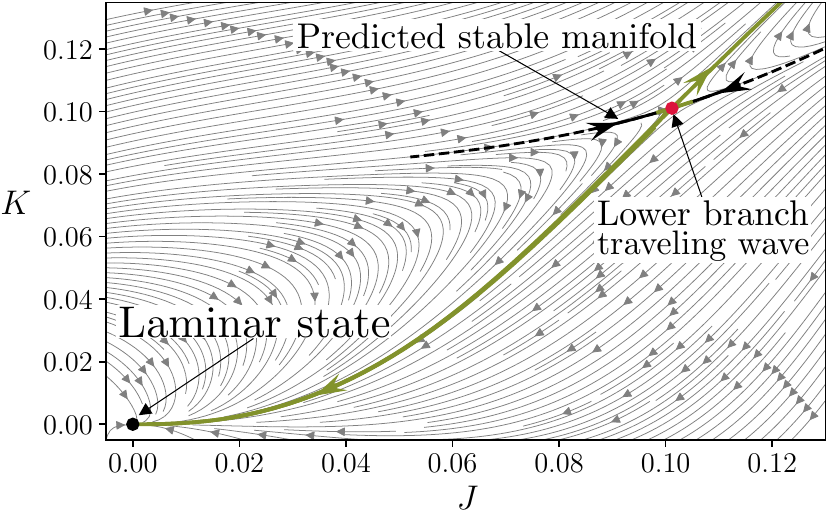}     
     \caption{Two-dimensional reduced-order model obtained by restricting the dynamics to the two-dimensional mixed-mode SSM of the lower-branch traveling wave. This manifold is parametrized by the variables $J$ and $K$ that define the reduced dynamics. Training trajectories are shown in green. The directory, including the data and the Jupyter notebook that generated this figure can be accessed at \url{https://cocalc.com/share/public_paths/db8c0a4e84051e09f620a97c03d6c18927f47ce0} }
     \label{fig:fig4}
 \end{figure}

In Fig. \ref{fig:fig4} we show the SSM-reduced model vector field $\mathbf{f}$ on the $J-K$ plane. We also show the approximation of the stable manifold of the lower-branch traveling wave, obtained by integrating initial conditions of the form
\begin{equation}
\begin{pmatrix}
    J_{LB} \\ K_{LB} 
\end{pmatrix} \pm 10^{-6} \mathbf{v}
\end{equation}
backward in time, where $\mathbf{v}\in \mathbb{R}^2$ is the eigenvector of $D\mathbf{f}(J_{LB}, K_{LB})$ associated with the negative eigenvalue. In Fig. \ref{fig:fig4}, the predicted stable manifold matches the edge trajectory well, even outside the domain of the training data for $J>J_{LB}$. 

\begin{figure}
     \includegraphics[width = \textwidth]{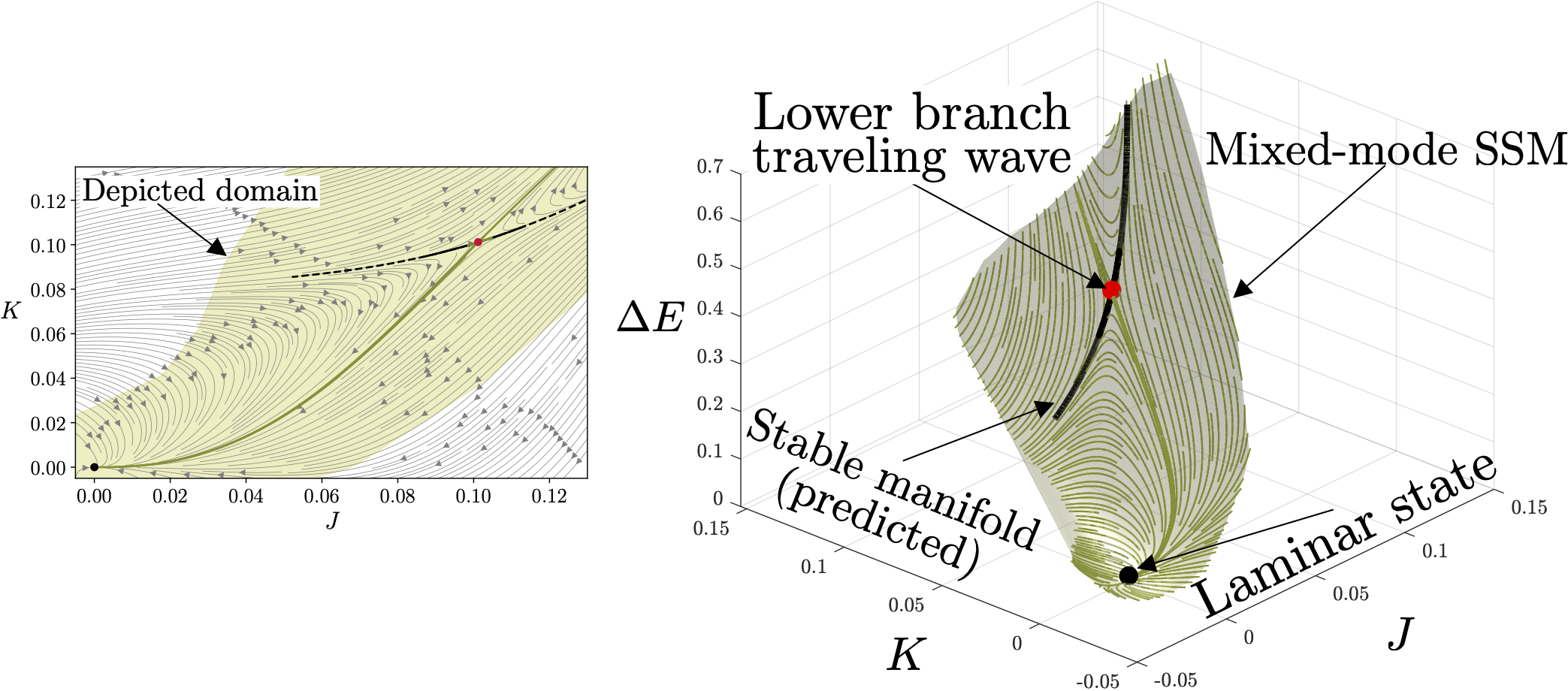}     
     \caption{Visualization of the reduced-order model and the geometry of the mixed-mode SSM carrying the model flow. The right panel shows the mixed-mode SSM (green surface) in the $(J,K,\Delta E)$ plane obtained as the image of the domain shown in the left panel  under the parametrization $\mathbf{W}$. The relative energy $\Delta E$ is defined as \eqref{eq:rele}. The fixed points, the base state and the lower-branch traveling wave are indicated with colored dots. The dashed line denotes the predicted stable manifold of the lower-branch traveling wave in the SSM-reduced model. The directory, including the data and the Jupyter notebook that generated this figure can be accessed at \url{https://cocalc.com/share/public_paths/f73549a1c2a577925859b24b039047cd5a05b1cd}}
     \label{fig:fig5}
 \end{figure}

Having computed the parametrization $\mathbf{W}$, we can also display the global shape of the mixed-mode SSM. In Fig. \ref{fig:fig5}, we highlight the domain in the $J-K$ plane, over which we visualize the SSM by plotting its relative kinetic energy,
\begin{equation}
\label{eq:rele}
    \Delta E = \frac{1}{2}\left \Vert \mathbf{u}_{HP} - \mathbf{W}(J,K) \right \Vert^2,
\end{equation}
in the right panel of Fig. \ref{fig:fig5}. 

\subsection{Capturing the edge of chaos in the reduced-order model}
\begin{figure}
     \centering
     \includegraphics[width = .75\textwidth]{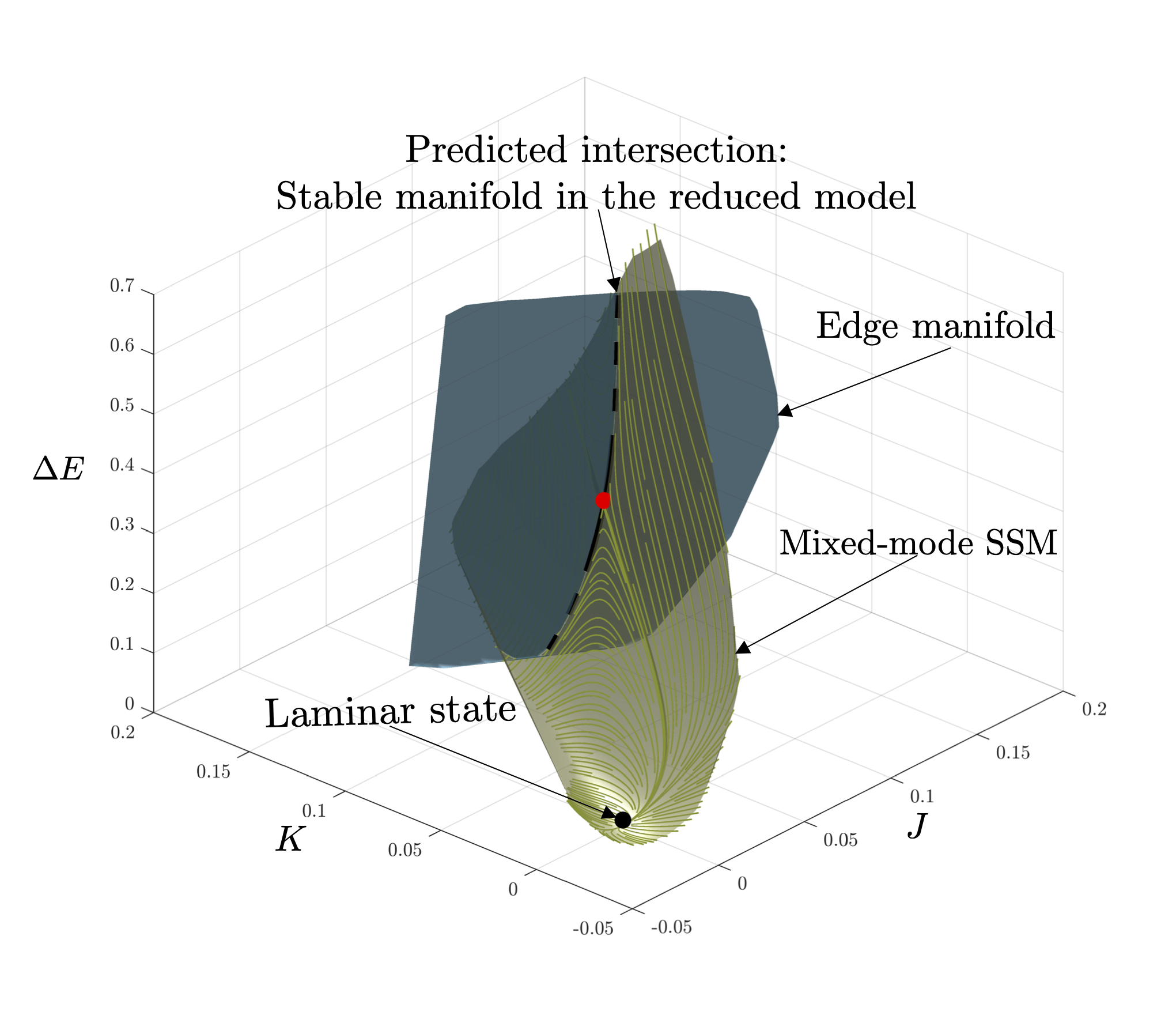}     
     \caption{Mixed-mode SSM and the edge manifold shown in the ($J,K,\Delta E$) space. The edge manifold is sketched as a two-dimensional surface in the three-dimensional space that intersects the edge manifold along the stable manifold of the lower-branch traveling wave in the reduced-order model (dashed black line). }
     \label{fig:fig6}
 \end{figure}

The mixed-mode SSM shown in Fig. \ref{fig:fig6} is two-dimensional, and the edge manifold ( i.e., the stable manifold of the lower-branch traveling wave) has codimension one. Adding up their dimensions results in $d+1$, where $d$ is the dimension of the phase space of \eqref{eq:symmetry_reduced_phasespace}. Therefore, these two manifolds generically intersect transversely along a one-dimensional curve. This implies that the intersection is robust under small perturbations to system \eqref{eq:ns_dynsys}. 

The generically expected transverse intersection of the SSM with the edge manifold is displayed in Fig. \ref{fig:fig6}, where the mixed-mode SSM is the same as in Fig. \ref{fig:fig5} and a schematic representation of the edge manifold is added. In the three-dimensional $(J, K, \Delta E)$ space, both the edge manifold and the mixed-mode SSM appear as two-dimensional surfaces, even though in the full phase space the edge manifold is a much higher dimensional object.

From the spectrum of the lower-branch traveling wave in Fig. \ref{fig:fig1}, we  conclude that its slowest eigenvalue with negative real part is real. Therefore, the stable manifold of this traveling wave has a one-dimensional slow SSM tangent to the eigenspace of that real eigenvalue. This SSM coincides with the stable manifold of the lower-branch traveling wave in the reduced-order model, and hence must be the line of intersection shown in dashed lines in Fig. \ref{fig:fig6}. 

\begin{figure}
     \centering
     \includegraphics[width = .99\textwidth]{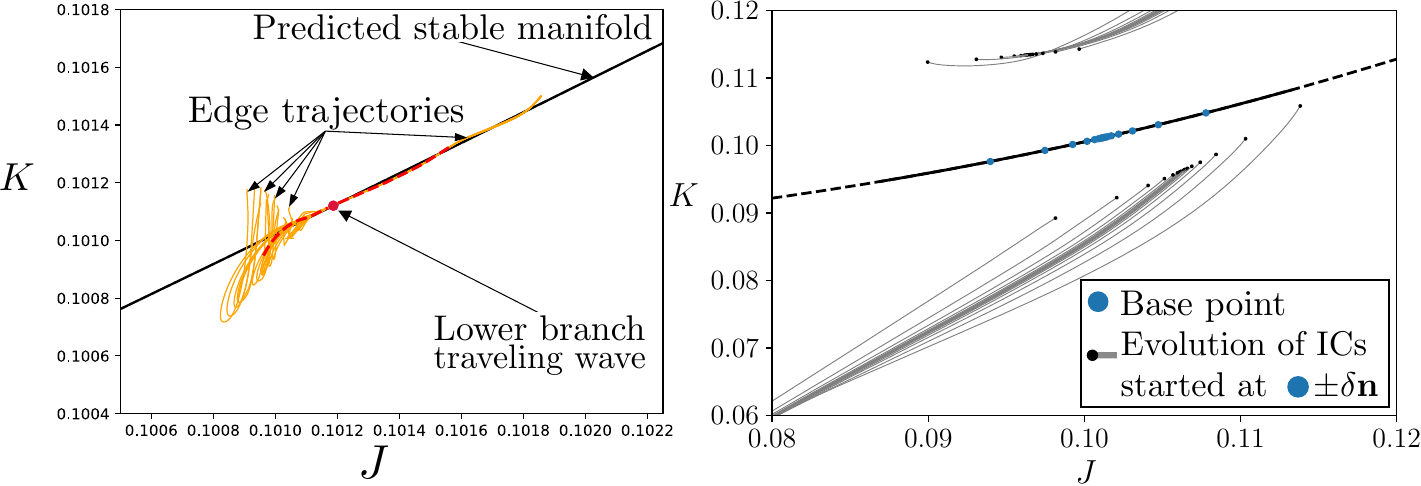}     
     \caption{Left panel: Edge trajectories constructed using the bisection algorithm in the full-order model approaching the lower-branch traveling wave displayed in the reduced-phase space. A single representative member of this ensemble of trajectories is shown with a red dashed line to guide the eye. The stable manifold of the lower-branch traveling wave predicted from the reduced-order model is shown in black. In the right panel, trajectories of the full-order  model are initialized by placing initial conditions on both sides of the predicted edge in the reduced-order model. }
     \label{fig:fig7}
 \end{figure}

Since the intersection of the edge manifold and the mixed-mode SSM is a curve, it will not capture the dynamics within the whole edge manifold. Nevertheless, this intersection identifies a clear footprint of the edge manifold in the SSM-reduced model. To demonstrate this, we construct 10 trajectories constrained to the edge manifold using the bisection algorithm of \cite{itano_DynamicsBurstingProcess2001}. These trajectories converge to the lower-branch traveling wave, which is a relative attractor within the edge. In the reduced phase space $(J,K)$, these trajectories clearly approach the lower-branch traveling wave along the predicted stable manifold, as shown in the left panel of Fig. \ref{fig:fig7}. 

The distinguishing property of the edge manifold is that it divides the phase space locally around the lower-branch traveling wave. We show that its footprint in the reduced-order model also has this dividing role by initializing a set of trajectories in the reduced phase space at a fixed distance $\delta=0.01$ from the stable manifold. Initial conditions are then prepared using the parametrization $\mathbf{W}$ and are supplied to the full-order model \texttt{Openpipeflow}. The right panel of Fig. \ref{fig:fig7} shows that the initial conditions prepared in this way are clearly separated in both the SSM-reduced model  and the full-order  model. 

\begin{figure}
     \centering
     \includegraphics[width = .99\textwidth]{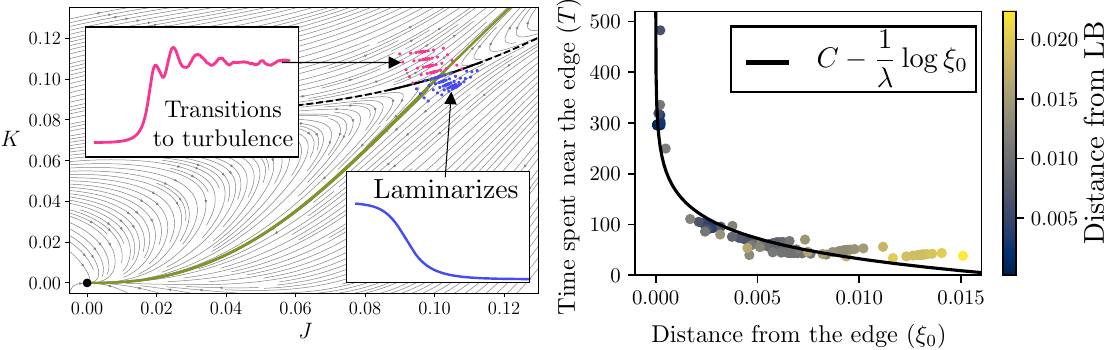}     
     \caption{Left panel: Phase portrait of the SSM-reduced model with a grid of test trajectories. Initial conditions are placed in a grid around the lower-branch traveling wave and are integrated forward in time using the full-order  model. The initial conditions are colored according to their long-time behavior, as seen in the insets. The predicted edge manifold footprint is shown in black. Right panel:time spent by these trajectories near the edge as a function of their initial distance from the edge measured in the SSM-reduced phase space. This time is defined as the time needed to reach a distance of $D_{\max}  = 0.1$ from the lower-branch traveling wave. A least-squares fit of the form $T=C - \frac{1}{\lambda}\log \xi_0$ is also shown with $C=-234 \pm 14 $ and $\lambda = 0.0175 \pm 0.0005$. }
     \label{fig:fig8}
 \end{figure}

In the left panel of Fig. \ref{fig:fig8}, we show a grid of initial conditions in a rectangle around the lower-branch traveling wave in the reduced-order model. The initial conditions for the full-order model are prepared using the identified parametrization $\mathbf{W}$ and are integrated forward in time. Different colors indicate the different long-time behaviors of the grid of initial conditions. The stable manifold of the reduced-order model clearly separates the trajectories of the full-order  model, as expected.

Another indication that this separation is indeed caused by the saddle-type edge state is the slowdown of trajectories, which is already present in a linear system. Consider a linear system with a saddle-type fixed point that has a one-dimensional unstable manifold. In diagonal form, the dynamics along the unstable subspace are
\begin{equation}
    \dot{\xi} = \lambda \xi, \text{ for } \lambda > 0,
\end{equation}
 while for the remaining coordinates we have
\begin{equation}
\dot{\eta}_i=-\kappa_i \eta_i, \text{ for } \kappa_i>0, i=1, ..., d-1.
\end{equation} For sufficiently long times, the distance of a trajectory from the saddle is well approximated by its $\xi$ coordinate component, i.e. the distance $\delta(t)$ can be written as $\delta(t)=\xi_0e^{\lambda t}$. Therefore, the time $T$ it takes for a trajectory to leave the $\delta_{\max}$-neighborhood of the saddle satisfies $\delta_{\max}=\xi_0 e^{\lambda T}$, or, equivalently
\begin{equation}
\label{eq:scaling}
T = \frac{1}{\lambda}\log \delta_{\max} - \frac{1}{\lambda}\log \xi_0,
\end{equation}
where $\xi_0$ measures the initial distance from the stable subspace. 

To demonstrate the slowdown, starting from the initial conditions shown in the left panel of Fig. 
\ref{fig:fig8}, we record the time needed for the full-order model trajectory to develop a distance of $\delta_{\max} =0.1$ from the lower-branch traveling wave. The right panel of Fig. \ref{fig:fig8} shows the dependence of this time on the initial distance from the edge manifold in the reduced model. This relationship can be well described by a logarithmic function of the form \eqref{eq:scaling}, which would hold for a linear system. Therefore, we expect that $T=C -\frac{1}{\lambda}\log \xi_0$ is satisfied approximately for some constants $C$ and $\lambda$. 
We obtain $\lambda = 0.0175$ from a least squares fit to the data, which reasonably matches the true unstable eigenvalue of the lower-branch traveling wave, $\lambda^{(+)}$.

Although the one-dimensional footprint of the edge manifold cannot act as a global barrier in the phase space, it is still an influential curve for trajectories near the lower-branch traveling wave. Since the mixed-mode SSM is constructed to be tangent to the slowest dynamics, it is locally attracting. Therefore, one may project nearby trajectories onto the mixed-mode SSM to decide their long-term dynamics: if the projection lies above the footprint of the edge manifold within the SSM, then the trajectory is expected to become turbulent. Otherwise, the trajectory is expected to quickly laminarize. As one moves away from the vicinity of the lower-branch, this correspondence will gradually break down due to the overall complicated shape of the edge (\cite{schneider_TurbulenceTransitionEdge2007, mellibovsky_2009}). 
 \subsection{Parameter dependent SSM-reduced models}
The spectrum of the lower-branch traveling wave changes smoothly as the Reynolds number is varied, therefore the slowest mixed-mode SSM remains two-dimensional for a wide range of Reynolds numbers. As in \cite{kaszas_DynamicsbasedMachineLearning2022}, we may then seek a parametrization of the Reynolds number-dependent family of mixed-mode SSMs. Similarly, the reduced dynamics on the family of manifolds then also depend on the Reynolds number. Formally, this means that we seek $\mathbf{W}$ and $\mathbf{f}$ in a Reynolds number-dependent form 

\begin{equation}
\label{eq:parametrization_re}
    \mathbf{u}(r,\varphi,z) = \mathbf{W}(J,K, \text{Re}) = \mathbf{u}_{HP}(r) + \sum_{l=0}^{N_p}\sum_{1\leq n+m\leq M_p}\mathbf{w}_{nml}(r, \varphi,z)K^mJ^{n}\text{Re}^l
\end{equation}
and 
\begin{equation}
\label{eq:reddynamics_re}
\frac{d}{dt}\begin{pmatrix}J  \\ K \end{pmatrix} = \mathbf{f}_M(J,K, \text{Re})= \sum_{l=0}^{N_p}\sum_{1\leq n+m\leq M_r} \begin{pmatrix} R^{(J)}_{nml} K^m J^n \text{Re}^l \\ R^{(K)}_{nm} K^m J^n \text{Re}^l\end{pmatrix}. \\
\end{equation}
The general dependence of these expressions on the Reynolds number is restricted by the requirement that the laminar state must be a fixed point at $J=K=0$ for all values of the Reynolds number, therefore $\mathbf{w}_{00l} =0$ and $R_{00l}=0$ must hold for all $l$. 

\begin{figure}
     \centering
     \includegraphics[width = .75\textwidth]{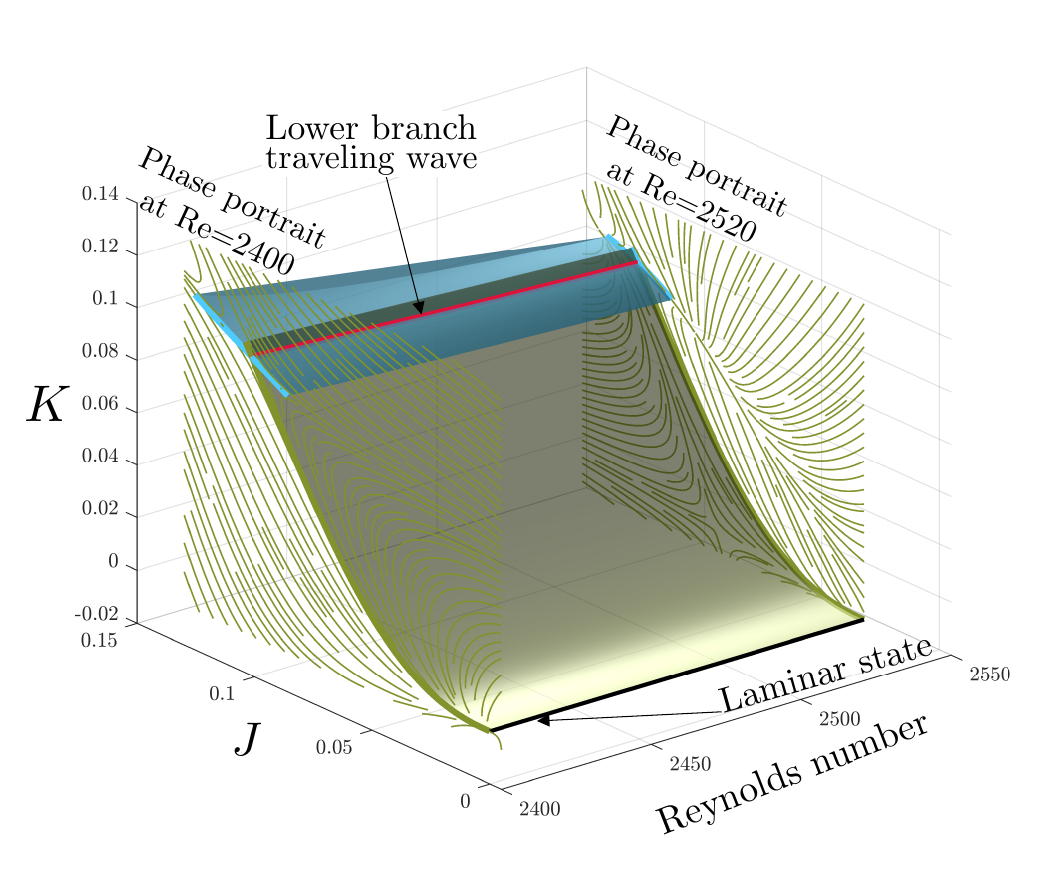}     
     \caption{Family of SSM-reduced models parametrized by the Reynolds number. Phase portraits at the sections Re = $2,400$ and Re = $2,520$ are shown, along with the lower-branch traveling wave and its stable and unstable manifolds computed from the SSM-reduced models. }
     \label{fig:fig9}
 \end{figure}

To construct a parameter-dependent SSM-reduced model, we select the range ~${\text{Re}\in[2400, 2520]}$, where the slowest eigenvalues of the lower-branch traveling wave do not change considerably and hence the mixed-mode SSMs are only expected to show minor variation. In this case, we can take a linear approximation for the Reynolds-number dependence in \eqref{eq:parametrization_re} and \eqref{eq:reddynamics_re}, i.e. $N_p=1$. We use three sets of training trajectories initialized at $\text{Re} = 2,400; 2,550; 2,520$. These are prepared in the same way as described in the case of $\text{Re}=2,400$. Figure \ref{fig:fig9} shows this parametrized family of reduced dynamics in the ($\text{Re}, J, K$) space. We find that the phase portraits of the reduced models are qualitatively the same for all Reynolds numbers in this range. 

\section{Conclusion}
\label{sec:conclusion}
We have constructed and tested invariant manifold-based reduced-order models for a transitional pipe flow at $Re=2,400$.
Specifically, we have computed a mixed-mode SSM, as recently defined by \cite{haller_NonlinearModelReduction2023}, to capture the slowest dynamics characteristic of the lower-branch traveling wave in the pipe flow. Following the numerical setup of \cite{willis_RevealingStateSpace2013}, we used a symmetry-restricted version of the dynamics, resulting in the lower-branch traveling wave coinciding with the edge state. In addition, we used the method of slices to factor out the physically irrelevant directions for the time evolution of the trajectories. We believe that the present study is the first example of applying SSM-based reduced-order models to systems exhibiting symmetry. We have used a Python-based implementation of the data-driven SSM-reducion method, \texttt{SSMLearnPy} (\cite{cenedese_SSMLearnPy2023}). To make the dataset accessible, we provide it in a compressed format by performing linear dimensionality reduction first (principal component analysis) on the full dataset to reduce its size. We stress, however, that the results presented here were all obtained using the complete dataset.

The intersection of the mixed-mode SSM with the stable manifold of the edge state revealed a structurally stable curve in phase space, the slowest submanifold within the edge manifold. We have demonstrated by direct numerical simulation that the extracted curve, the one-dimensional footprint of the edge manifold, already displays edge-like characteristics: it separates laminarizing and turbulent trajectories in the phase space. This illustrates that the identified structure can be used to predict whether a given initial condition develops turbulence or simply laminarizes. We have also constructed parameter-dependent SSM-reduced models that remain valid over a range of Reynolds numbers.

For simplicity, we carried out the calculations following an additional step of symmetry reduction through the method of slices. To obtain structures in the full phase space, one simply takes the direct product of the identified structures with the group orbit of streamwise translations. For example, the mixed-mode SSM of Fig. \ref{fig:fig6} becomes a three-dimensional structure and its intersection with the edge manifold becomes a cylinder in the full phase space.
Alternatively, using the non-sliced simulation data, one may look for the mixed-mode SSM as a 3-dimensional manifold. 

In this study, we utilized the small domain size and the additional shift-reflect symmetry and the 2-fold azimuthal rotation symmetry restrictions. These assumptions made it possible to conclude that the edge manifold is locally of codimension one, enabling the easy characterization of its slowest submanifold as the intersection with the mixed-mode SSM. Although these assumptions fail for more general pipe flows, a similar approach could be employed even in longer pipes (\cite{avila_TransientNatureLocalized2010, avila_StreamwiseLocalizedSolutionsOnset2013a}) without imposing the symmetries. By generating trajectories restricted to the edge of chaos using the bisection method (\cite{itano_DynamicsBurstingProcess2001}), one could find the most influential submanifolds within the edge along with the dynamics restricted onto them. This would, however, require constructing invariant manifolds of considerably higher dimensions which brings additional challenges.

\section*{Acknowledgements}
We are grateful to Florian Kogelbauer, Ashley Willis, and Mingwu Li for several insightful conversations. The numerical simulations were performed on the Euler cluster operated by the High Performance Computing group at ETH Zürich.
We acknowledge support from the Turbulent Superstructures Program
(SPP1881) of the German National Science Foundation (DFG).

\section*{Data availability}
The code and data (in compressed format) used in the paper is available in the repository \url{https://github.com/haller-group/SSMLearnPy}. The complete dataset is available from the authors upon request.

\section*{Declaration of Interests}
The authors report no conflict of interest.
\section*{Appendix: Resonance conditions}
The mixed-mode SSM of the lower-branch traveling wave, as constructed by \cite{haller_NonlinearModelReduction2023}, exists under the conditions of the linearization theorem of \cite{sternberg_StructureLocalHomeomorphisms1958}. These require that the eigenvalues of the Jacobian of \eqref{eq:ns_dynsys} at the lower-branch traveling wave be nonresonant. Specifically, denoting the eigenvalues as $\lambda_i$ and ordering them according to descending real part, we require
\begin{equation}
\label{eq:resonance}
    \lambda_j \neq \sum_{i=1}^d m_i \lambda_i, \quad m_i\in \mathbb{N} \quad \sum_{i=1}^d m_i\geq2.
\end{equation}
We check that the conditions \eqref{eq:resonance} hold for a finite subset of the spectrum of the lower-branch traveling wave by constructing all admissible integer linear combinations of the leading $23$ eigenvalues shown in the right panel of Fig. \ref{fig:appendix}. To select the eigenvalues nearest to resonance, for every eigenvalue $\lambda_j$, we define
\begin{equation}
    D(\lambda_j) = \min_{m_i} \left \vert\lambda_j - \sum_{i=1}^d m_i\lambda_i \right \vert,
\end{equation}
which measures how close $\lambda_j$ is to being resonant. We plot the relative measure of resonance-closeness, $D(\lambda)/|\lambda|$ for the leading eigenvalues in the left panel of Fig. \ref{fig:appendix}.
\begin{figure}
\label{fig:appendix}
    \centering
    \includegraphics[width = \textwidth]{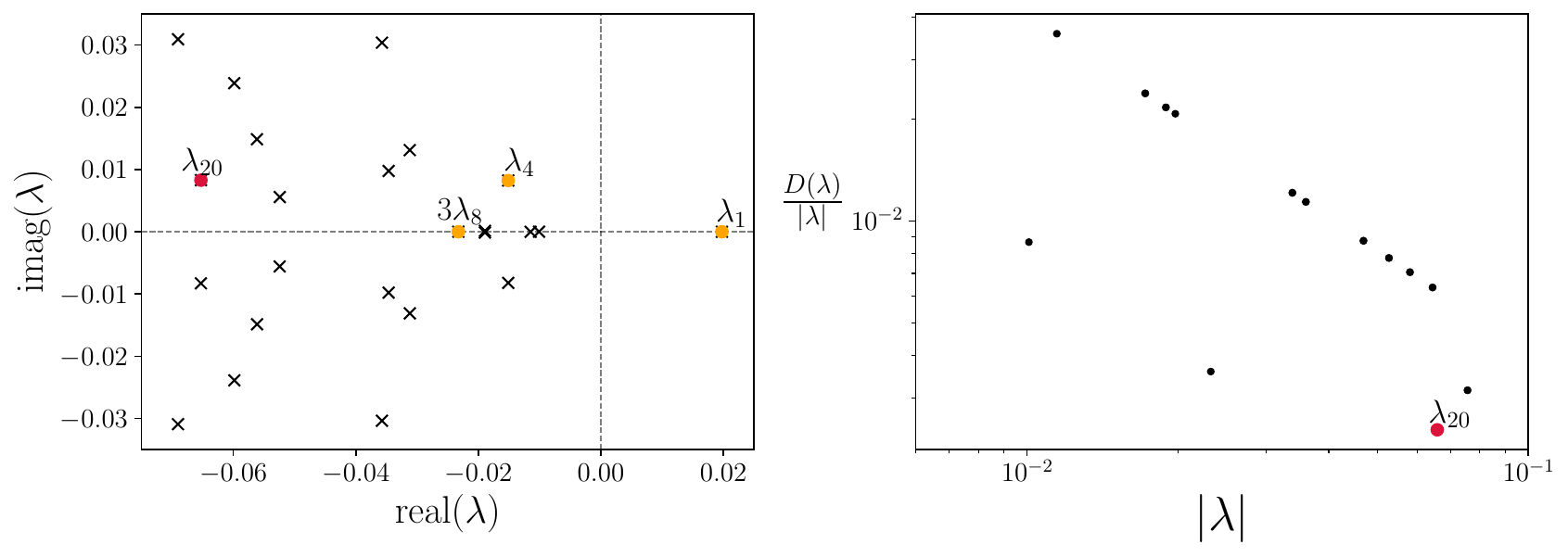}
    \caption{Left panel: Leading eigenvalues of the lower-branch traveling wave restricted to the $\mathcal{S}-$invariant subspace. Right panel: relative measure of resonance-closeness, $D(\lambda)/|\lambda|$ as a function of $|\lambda|$. The eigenvalue closest to being resonant, $\lambda_{20}$, is marked with a red point. The three eigenvalues generating this close resonance, $\lambda_1$, $\lambda_4$, and $\lambda_8$ are marked with orange points. The directory, including the data and the Jupyter notebook that generated this figure can be accessed at \url{https://cocalc.com/share/public_paths/186d749169aa46fdb63d21756f1325b50964ac2e}}
    \label{fig:appendix}
\end{figure}
Although no exact resonance is seen in the spectrum, near-resonances of up to $0.1\%$ coincidence can be seen. The nearest resonance is highlighted in Fig. \ref{fig:appendix} as $\lambda_{20}$, whose value is
\begin{equation}
    \lambda_{20}=-0.06529 + 0.00829i.
\end{equation}
With a linear combination of $\lambda_1$, $\lambda_4$, and $\lambda_8$, we get
\begin{equation}
    \lambda_{1}+\lambda_4 + 3\lambda_8=-0.06514 + 0.00822i,
\end{equation}
which yields $D(\lambda)/|\lambda|\approx 0.2 \%$. Although close resonances of similar severity may occur in the rest of the spectrum, we see no indication of exact resonances, hence the results of \cite{haller_NonlinearModelReduction2023} should apply.


\end{document}